\documentclass[12pt]{article}
\usepackage{aasms4,psfig}

\newcommand{\eg}{{\it e.g.\ }}
\newcommand{\etal}{{\it et al.\ }}

\begin{document}

\title{ON THE ORIGIN OF THE MeV GAMMA-RAY BACKGROUND}
\author{F. W. Stecker}
\affil{Laboratory for High Energy Astrophysics, NASA Goddard Space 
Flight Center, Greenbelt, MD 20771, U.S.A.}
\authoremail{stecker@lheavx.gsfc.nasa.gov}

\author{M. H. Salamon}
\affil{Physics Department, University of Utah, Salt Lake City, UT 84112, U.S.A.}
\authoremail{salamon@physics.utah.edu}

\author{C. Done}
\affil{Physics Department, University of Durham, Durham, UK}
\authoremail{chris.done@durham.ac.uk}

\begin{abstract}

In this paper, we suggest a new hypothesis for explaining the spectrum of the 
extragalactic MeV $\gamma$-ray background as observed by COMPTEL and SMM. 
We propose that both the flux level and spectrum can be accounted for as a 
superposition of non-thermal MeV tails in the spectra of Seyfert galaxies and 
other AGN. Although present detectors are not sensitive enough to obtain MeV 
data from individual extragalactic sources, indirect evidence in support of 
our hypothesis is found in OSSE and COMPTEL observations of the galactic 
black hole candidate Cygnus X-1. 

\end{abstract}


\section{Introduction}

There has been much recent progress over the last 10 years in understanding the
origins of the high energy cosmic background radiation. It now seems almost
certain that the bulk of the hard X--ray background from 2--200 keV is made 
from obscured radio--quiet AGN (Madau, Ghisellini \& Fabian 1994; 
Comastri \etal 1995; Gilli, Risalti \& Salvati 1999), 
while at high 
energies (30 MeV -- 100 GeV) the beamed radio--loud AGN (blazars) dominate 
(Stecker \& Salamon 1996; Zdziarski 1996; Sreekumar \etal 1998). However, 
between the radio--quiet AGN rollover at
$\sim 100 $ keV, and the low energy break in the spectrum of radio--loud AGN at
$\sim 10$ MeV, there is a substantial background component detected by COMPTEL
and SMM in the 200 keV -- 3 MeV range which is not accounted for in these
models. Some of this is produced by type Ia supernovae (Zdziarski 1996), 
but the latest calculations show that there is still a marked discrepancy 
(by about a factor 2) between the observed background and current best 
estimates of the supernovae contribution (Watanabe et al 1999). 
It also appears that blazars 
will not account for the MeV background because their spectra generally have a 
break at an energy of about 10 MeV (McNaron-Brown \etal 1995). The small
population of possible ``MeV blazars'' will also not account for the MeV 
background. In fact, even if we accept that unresolved blazars 
account for the
extragalactic background radiation at energies above 30 MeV (Stecker \& Salamon
1996), essentially all of these blazars would have to be ``MeV blazars'' as 
well in order to account for the background flux level at MeV energies,
contrary to the observational evidence (McNaron-Brown \etal 1995).

In this paper we propose a new hypothesis to account for the
extragalactic background in the MeV region as derived from two
independent analyses of the COMPTEL data from the Compton Gamma Ray
Observatory satellite (Kappadath, \etal 1996; Sreekumar, Stecker and
Kappadath 1997; Weidenspointner 1999).  Our hypothesis is based on an
analogy between the galactic black hole candidate Cyg X--1 and active
galactic nuclei (AGN), which are generally believed to be powered by
supermassive black holes. We use this analogy to extend the
observation of a nonthermal MeV ``tail'' in the Cyg X--1 spectrum to
hypothesize that such nonthermal tails exist in extragalactic AGN
spectra, even though past and present gamma-ray detectors could not
observe such tails at the flux levels expected. We will then argue
that a superposition of unresolved AGN with Cyg X--1 type spectra,
such as has been shown to reasonably account for the X-ray background
(\eg Gilli, Risaliti, and Salvati 1999) can also account for the shape
and flux level of the MeV background deduced from the COMPTEL data.

In this regard, it is relevant to note very recent obervations of Seyfert 
galaxies with flat spectrum radio nuclei using the VLBA have shown that these
sources are emitting non-thermal radiation from central core regions with 
sizes $\sim$ 0.05 to 0.2 pc (Mundell, Wilson, Ulvestad \& Roy 1999). Such 
cores may also be the source of non-thermal MeV emission.

\section{The Cyg X--1 Epitome}

Cyg X--1 in its low/hard state has a
spectrum dominated by a power law component which rolls over at $\sim 200$ keV.
This is well fit by a model involving a thermal population of hot electrons
which Compton upscatter soft seed photons from the accretion disk
(e.g. Gierlinski et al 1997). However, recent COMPTEL observations show a small
hard tail of emission, extending out to MeV energies (McConnell \etal 1997).
An explanation that has been suggested to account for this hard tail would be 
that the electron
distribution is not completely thermalized (Poutanen \& Coppi 1998). This is
physically reasonable since the thermalization timescales for the electrons can
be rather slower than the other timescales in these systems (e.g. Coppi
1999). The overall 2 keV -- 5 MeV spectrum of Cyg X--1 can then be modelled if
$90$ per cent of the power goes into a $\sim 100$ keV thermal electron
distribution, while the remaining $\sim 10$ per cent is in the form of a
non--thermal tail (Poutanen \& Coppi 1998).

It is well known that the low/hard state spectrum of Cyg X--1 and other 
galactic black hole candidates
bear a remarkable similarity to that from radio--quiet AGN
(see e.g. the review by Poutanen 1998), plausibly because
both involve the same physical processes of disk accretion onto a black
hole. Thus we expect a similar hard tail to be present in Seyfert galaxies
(both type 1 and type 2). In
Cyg X--1 this tail begins 
roughly an order of magnitude below the peak in the hard X--ray
spectrum. Such a tail could not be detected in an individual AGN using current
instrumentation, but we will show that a superposition of such tails in the
spectra of AGN would account for the reported 
MeV background spectrum and flux.

\section{A Further Component to the MeV background from Seyferts ?}

Galactic black hole candidate sources are known to make spectral
transitions to a high/soft state at accretion
rates of greater than 10 per cent of the Eddington mass accretion rate. 
In this state the
spectrum is dominated by a thermal component at $\sim 1$ keV (presumably
corresponding to emission from the accretion disk), but also shows a steep
non--thermal hard X--ray tail which extends past 511 keV (Grove et al 1998;
Gierlinski et al 1999). It is not yet known where this power law breaks
(Grove et al 1998), but it is plausible that this also extends to MeV
energies, as seems to be indicated by COMPTEL data from Cyg X--1
(Poutanen \& Coppi 1998; Gierlinski et al 1999). There is a class of Seyfert
galaxies, the Narrow Line Seyfert 1's Osterbrock \& Pogge 1985; Boroson \&
Green 1992), which are thought to be the AGN analogue
of these high mass accretion rate systems (Pounds, Done \& Osborne
1995). These comprise about 10 per cent of Seyferts (Boroson \& Green
1992), so would also contribute to an extragalactic MeV background
if these truely are comparable to the soft/high state galactic
black holes in having a steep unbroken power law 
spectrum extending beyond 511 keV.

\section{An Illustrative Spectrum}

We will now estimate contributions that AGN power-law MeV tails would make
to the X-ray/MeV background were these to be universal components in AGN
spectra.  Because our main concern is with the MeV component of the
diffuse background, we restrict our
calculation of the extragalactic background spectrum 
to energies in the range 100 keV - 10 MeV.
Our calculation of the X-ray background (XRB) follows that
of Pompilio, La Franca \& Matt (1999) (PLM), to which we refer the reader for
details.  Briefly, the XRB is assumed to be comprised of the summed emission
of unresolved AGN, which fall into two types: AGN1 and AGN2.  In the standard
unification scheme, the spectral differences between the two are due to the
orientation of the AGN molecular torus relative to our line of sight.  For
AGN1, there is no obscuration of the nucleus by the torus along our line
of sight, while for AGN2 the nuclear spectrum is both attenutated and altered
by photoelectric absorption and Compton scattering within the torus.

Following Comastri \etal (1995), PLM adopt the following for the AGN1
source luminosity $l(E)$ in units of keV s$^{-1}$keV$^{-1}$
\begin{equation}
l(E)\propto  \left\{ \begin{array}{ll}
E^{-1.3} & E<1.5 \\
E^{-0.9}e^{-E/400}+ r(E) & E>1.5
\end{array} \right.,
\end{equation}
where $E$ is in keV, and $r(E)$ is a Compton reflection component 
of nuclear emission off the surrounding gas and dust.  The normalization
coefficient is determined by requiring $\int l(E)\,dE = L$, where
$L$ is the
AGN luminosity in the energy band 0.3-3.5 keV at the source.
To this we have added a non-thermal
power-law component with a soft cutoff at energies below the XRB peak
of $\sim30$ keV whose amplitude and spectral index are variable parameters.
We neglect the reflection component, which
is essentially compensated for in our energy region 
by normalizing the AGN1 spectrum to the AGN luminosity $L$.  Integration
of
\begin{equation}
l(E)= \kappa L \left\{ \begin{array}{ll}
(E/1.5)^{-1.3} & E<1.5 \\
(E/1.5)^{-0.9}e^{-E/400}+\eta(E/1.5)^{-\alpha}{\cal C}(E) & E>1.5
\end{array} \right.
\end{equation}
over $0.3<E<3.5$ keV determines the normalization constant $\kappa$.  
The power-law tail is assumed
to be cut off below 30 keV by an arbitrary cutoff function ${\cal C}(E)$.

The spectra of AGN2 are modifed by the intervening molecular tori.  
The ratio $R(z)$ of AGN2 to AGN1 sources is taken from PLM, and is
approximately 5 at redshift $z=0$.  The distribution of 
torus thicknesses through which the AGN2 emissions pass is taken from
Risaliti, Maiolino \& Salvati (1999) (their Table 3).  Above
100 keV photoelectric absorption is negligible compared to Compton
scattering (Morrison \& McCammon, 1983), so 
we consider only the latter in calculating
the mean transmission coefficient $T(E)$ for AGN2 spectra.

The AGN1 X-ray luminosity function (XLF) $\Phi(L,z)$, following PLM, is taken
to be separable, $\Phi(L,z)=\Phi_{0}(L)f(z)$, where $f(z)$ is the
evolution factor, and $\Phi_{0}(L)$ is
the current XLF.  Integrating over the XLF gives the intensity of the 
diffuse background
$I(E)$ in units of keV cm$^{-2}$s$^{-1}$
sr$^{-1}$keV$^{-1}$ 
\begin{equation}
I(E)=\frac{c}{4\pi H_{0}}\int_{0}^{z_{\rm max}}
dz\,\frac{f(z)l\left[E(1+z)\right]\left\{
1+R(z)T\left[E(1+z)\right]\right\}}
{(1+z)^{2}(1+2q_{0}z)^{1/2}}
\int_{L_{\rm min}}^{L_{\rm max}}dL\,L\Phi_{0}(L).
\end{equation}

Figure 1 shows the results of this calculation with two power-law
tail spectral indices, $\alpha=$1.2 and 1.4.  The amplitude
$\eta$ for these two cases is chosen to best fit the set of diffuse
background measurements; in both cases this results in the power-law
tail component being roughly an order of magnitude below the 30 keV
peak value of the XRB.
This is consistent (within the considerable uncertainties)
with the amplitude of the MeV tail of Cyg X--1 relative to its
hard-state peak. 

\section{Conclusion}

We have examined a new hypothesis for explaining the origin of the
extragalactic background radiation at MeV energies. Based on data from
the galactic black hole candidate Cygnus X-1, and assuming that radio
quiet active galactic nuclei, {\it i.e.} the Seyfert galaxies, contain
much more massive black holes at their cores, we assume that all such
black hole sources exhibit a high energy tail of the same magnitude
relative to the thermal emission as Cygnus X-1.  We show that by
making this assumption, we can account for the flux and spectrum of
the extragalactic MeV background as a superposition of emission from
Seyfert AGN.

\section{Acknowledgements}
We thank P. Sreekumar for use of his compilation of diffuse background
data (Sreekumar et al., 1998).

\newpage

\centerline{}
   \centerline{\psfig{figure=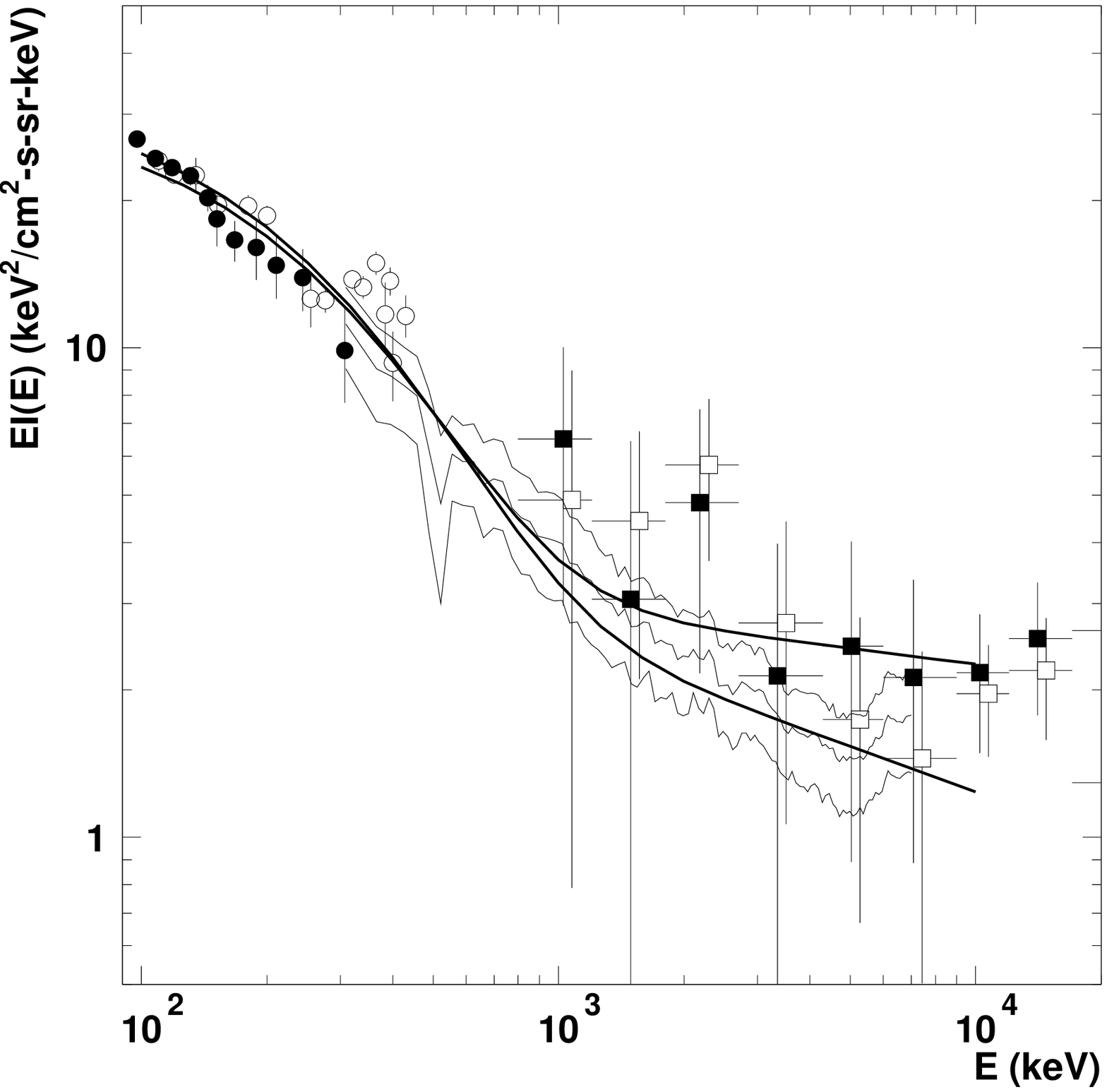,width=5.5in}}
\figcaption{Calculated diffuse X-ray and MeV gamma-ray background from
AGN1 and AGN2 sources to whose spectra a power-law tail has been
added.  The two solid lines correspond to power-law tails of energy
indices $\alpha=1.2$ and 1.4 at the source.  These are slightly
hardened after passage through the molecular tori due to the energy
dependence of the Compton cross section.  Also shown is data from the
GCRO/COMPTEL instrument (filled squares: Weidenspointner 1999; open squares
(slightly displaced in energy): Kappadath \etal 1996), the
Solar Maximum Mission (central thin line, and two 1$\sigma$ lines:
Watanabe \etal 1997 and 1999), and HEAO-A4 (filled circles: Kinzer
\etal 1997; open circles: Gruber \etal 1992).}


\begin{thebibliography}{99}

\bibitem[]{bor92} Boroson,T.A., Green R., 1992, ApJS., 80, 109

\bibitem[]{com95} Comastri,A., Setti,G., Zamorani,G. \& Hasinger,G. 1995,
A \& A 296, 1

\bibitem[]{cop99} Coppi, P. 1999, High Energy Processes in Accreting Black
       Holes, eds. J. Poutanen \& R. Svensson, ASP Conf. Ser. Vol. 161, 375 

\bibitem[]{gie99} Gierlinski, M., Zdziarski, A.A.,
Poutanen, J.,  Coppi, Paolo S. Ebisawa, K., Johnson, W.N., 1999,
MNRAS, 309, 496

\bibitem[]{gil99} Gilli, R., Risaliti, G. \& Salvati, M. 1999, 
A\&A., 347, 424

\bibitem[]{gro98} Grove, J.G., \etal 1998, ApJ 500, 899

\bibitem[]{gru92} Gruber, D.E. 1992, The X-ray Background, ed. X. Barcons
        \& A.C. Fabian (Cambridge: Cambridge Univ. Press) 44

\bibitem[]{kap96} Kappadath, S.C. \etal 1996, A\&AS 120, 619

\bibitem[]{kin97} Kinzer, R.L. \etal 1997, ApJ 475, 361

\bibitem[]{mad94} Madau, P. Ghisellini, G. \& Fabian, A.C. 1994, 
	MNRAS 270, L17

\bibitem[]{mcc97} McConnell, \etal. 1997, Adv. Space Res., 19, 25

\bibitem[]{mcn95} McNaron-Brown, K. \etal 1995, ApJ 451, 575

\bibitem[]{mor83} Morrison, R. \& McCammon, D. 1983, ApJ 270, 119

\bibitem[]{mun99} Mundell, C.G., Wilson, A.S., Ulvestad, 
	J.S. \& Roy, A.L. 1999, ApJ, in press.

\bibitem[]{ost85} Osterbrock, D.E. \& Pogge, R.W. 1985, ApJ 297, 166

\bibitem[]{pom99} Pompilio, F., La Franca, F., and Matt, G. 1999, 
A\&A, in press

\bibitem[]{pou95} Pounds, K.A.  Done, C. \& Osborne, J.P.  1995, MNRAS 277, L5

\bibitem[]{pou98} Poutanen, J., 1998, Theory of Black Hole Accretion
Disks, ed. M.A. Abramowicz, G. Bjornsson, and J.E. Pringle (Cambridge,
Cambridge University Press), 100

\bibitem[]{pou98} Poutanen, J. \& Coppi, P. 1998, Phys.Scripta, T77, 57

\bibitem[]{ris99} Risaliti, G., Maiolino, R., \& Salvati, M. 1999,
ApJ. 522, 157

\bibitem[]{sre97} Sreekumar, P. Stecker, F.W. \& Kappadath, 
	S.C. 1997, Proc. 4th Compton Symp., AIP No 410, 344

\bibitem[]{sre98} Sreekumar, P. \etal 1998, ApJ 494, 523

\bibitem[]{ste96} Stecker, F.W. \& Salamon, M.H. 1996, ApJ 464, 600

\bibitem[]{wei99} Weidenspointner, G. 1999, Ph.D. Thesis, 
	Technische Universisaet Muenchen

\bibitem[]{wat99} Watanabe K. \etal 1999, ApJ 516, 285

\bibitem[]{wat97} Watanabe, K., et al. 1997, AIP Conf. Proc. 410, 1223,
	eds. C.D. Dermer, M.S. Strickman, and J.D. Kurfess

\bibitem[]{zdz96} Zdziarski, A.A. 1996, MNRAS 281, L9

\end{thebibliography}
\end{document}